\documentclass[a4paper,12pt,titlepage]{article}
\usepackage{amsmath}
\usepackage{amsfonts}
\usepackage{graphicx}
\usepackage{scrextend}
\usepackage{authblk}
\usepackage{algorithmic}
\usepackage{algorithm}
\usepackage{amsthm}
\usepackage[numbers,comma,sort&compress]{natbib}
\usepackage{float}
\usepackage{lineno}
\usepackage{multirow}
\usepackage[utf8]{inputenc}

\renewcommand{\phi}{\varphi}
\renewcommand{\epsilon}{\varepsilon}

\newcommand{\mathscr}{\mathcal} % mathscr is not defined in this template, daniel switches it to mathcal

\def\squareforqed{\hbox{\rlap{$\sqcap$}$\sqcup$}}
\def\qed{\ifmmode\squareforqed\else{\unskip\nobreak\hfil
\penalty50\hskip1em\null\nobreak\hfil\squareforqed
\parfillskip=0pt\finalhyphendemerits=0\endgraf}\fi}

\newtheorem{implementation}{Algorithm}
\newtheorem{problem}{Problem}

\newtheorem{control-problem}{Control Problem}

\newcommand{\id}{{\mathrm{id}}\,}

\newcommand{\argmax}{{\mathrm{argmax}}}

\newcommand{\ignore}[1]{}

\newcommand{\R}{\mathbb{R}}

\begin{document}

\renewcommand{\thefootnote}{\fnsymbol{footnote}}
\title{\vspace{-10ex}{On variational solutions for whole brain serial-section histology using the computational anatomy random orbit model}}
\author[1]{Brian C. Lee \thanks{Corresponding author: blee105@jhu.edu (BCL)}}
\author[1]{Daniel J. Tward}
\author[2]{Partha P. Mitra}
\author[1]{Michael I. Miller}
\affil[1]{Center for Imaging Science, Johns Hopkins University, Baltimore, MD, USA}
\affil[2]{Cold Spring Harbor Laboratory, Cold Spring Harbor, NY, USA}
\date{\vspace{-5ex}}
\maketitle
\renewcommand{\thefootnote}{\arabic{footnote}}

\begin{abstract}
This paper presents a variational framework for dense diffeomorphic atlas-mapping onto high-throughput histology stacks at the 20 $\mu$m meso-scale.
The observed sections are modelled as Gaussian random fields conditioned on a sequence of unknown section by section rigid motions and unknown diffeomorphic transformation of a three-dimensional atlas.
To regularize over the high-dimensionality of our parameter space (which is a product space of the rigid motion dimensions and the diffeomorphism dimensions), the histology stacks are modelled as arising from a first order Sobolev space smoothness prior. 
We show that the joint maximum a-posteriori, penalized-likelihood estimator of our high dimensional parameter space emerges as a joint optimization interleaving rigid motion estimation for histology restacking and large deformation diffeomorphic metric mapping to atlas coordinates. 
We show that joint optimization in this parameter space solves
the classical curvature non-identifiability of the histology stacking problem.  The algorithms are demonstrated on a collection of whole-brain histological image stacks from the Mouse Brain Architecture Project. 
\end{abstract}
\section*{Author Summary}
New developments in neural tracing techniques have motivated the widespread use of histology as a modality for exploring the circuitry of the brain. Automated mapping of pre-labeled atlases onto modern large datasets of histological imagery is a critical step for elucidating the brain's neural circuitry and shape. This task is challenging as histological sections are imaged independently and the reconstruction of the unsectioned volume is nontrivial. Typically, neuroanatomists use reference volumes of the same subject (e.g. MRI) to guide reconstruction. However, obtaining reference imagery is often non-standard, as in high-throughput animal models like mouse histology. Others have proposed using anatomical atlases as guides, but have not accounted for the intrinsic nonlinear shape difference from atlas to subject. 
Our method addresses these limitations by jointly optimizing reconstruction informed by an atlas simultaneously with the nonlinear change of coordinates that encapsulates anatomical variation. This accounts for intrinsic shape differences and enables rigorous, direct comparisons of atlas and subject coordinates.
Using simulations, we demonstrate that our method recovers the reconstruction parameters more accurately than atlas-free models and innately produces accurate segmentations from simultaneous atlas mapping. We also demonstrate our method on the Mouse Brain Architecture dataset, successfully mapping and reconstructing over 500 brains.
\section{INTRODUCTION}
\subsection{Mapping brain circuitry}
Recent advances in brain imaging \cite{Hagmann2008,Rubinov2010}, methods to label neurons \cite{Taniguchi2011}, and computational methods have brought about a new era of neuroanatomical research, with a focus on comprehensively mapping brain circuits \cite{Kasthuri2007}. Mapping whole-brain circuitry is important for three distinct reasons: scientific understanding of how the brain works, mechanistic understanding of neurological and neuropsychiatric disorders, and as a comparison point for artificial neural networks used in machine learning \cite{Mitra2014-circuit,Sporns2011}.

Circuit mapping is technique limited, and falls into three broad scales corresponding to distinct imaging modalities - indirect mapping at a macroscopic scale corresponding to MRI-based methods \cite{Hagmann2007}, and direct mapping at light (LM) and electron microscopic (EM)  scales. For MRI and LM data, atlas mapping is an important step in the analysis. Several approaches exist for gathering LM data at the whole brain level \cite{Osten2013,Okano2015,Papp2016}. 
For some of these approaches (two-photon serial block-face imaging, knife edge scanning microscopy and light sheet microscopy for cleared brains) two-dimensional (2D) optical sections are acquired in three-dimensional (3D) registry with each other, so that the only computational step required is 3D volumetric registration of the individual brain data set to a canonical atlas. However, for classical neurohistological approaches using tissue sectioning followed by histochemical processing, the 2D sections are gathered independently and each section can undergo an arbitrary rotation and translation compared to the block face. This may be considered a disadvantage of the classical neuroanatomical workflow, however the physical sectioning method followed by conventional histochemical analysis has certain important advantages. This allows for the full spectrum of histochemical stains, acquisition of physical sections for downstream molecular analyses,  and processing for larger brains (upto and including whole human brains). Therefore it is necessary to perform an intermediate 2D to 3D registration step, where the individually acquired 2D sections are mutually co-registered into a 3D volume. 

This paper develops a joint stack reconstruction and atlas mapping procedure that simultaneously restacks the 2D histology sections, applying a sequence of rigid motions to the sections, and estimates the diffeomorphic correspondence between the registered histology stack and the 3D atlas. We apply these algorithms to data sets from the Mouse Brain Architecture Project (MBAP), for which the experimental workflow generating the data utilizes a tape transfer technique \cite{mitra-tape-transfer-2015}, allowing for the sections to maintain geometrical rigidity within section and also allowing for physically disjoint components to maintain their spatial relations. The tape method ensures that the number of missing sections is minimal, with serial sections cut at a thickness of 20 $\mu$m and alternate sections subjected to Nissl staining alongside staining with histochemical or fluorescent label. These Nissl stained sections form the basis of alignment to a Nissl whole-brain reference atlas.

\subsection{Computational anatomy methods for brain histology}
The histological reconstruction problem has been explored by several groups previously. Malandain first described the ill-posedness of reconstructing 3D sections and object curvature without prior knowledge of the shape of the object \cite{Maladain2004}. 
Rigid transformations for stack reconstruction have been estimated via block-matching of histological sections in \cite{Ourselin2001}, 
with point information based on landmarks introduced to guide volume reconstruction \cite{Streicher1997}.
Dense external reference information such as MRI has been applied to guide reconstruction via registration of corresponding block-face photographs and for histology to MRI mapping \cite{Dauguet2007,Adler2014}.
Anatomical atlases have also been suggested as guides for reconstruction, but without accounting for the intrinsic nonlinear shape differences between an atlas and any given subject brain \cite{Qiu2011}.

The principal contribution of this work is to rigorously solve the problem when an external resource of identical geometry (such as an MRI of the same mouse) is not available, while accommodating for the innate anatomical variation from atlas to subject. The lack of a same-subject reference volume is often the standard in mouse brain histology and other large scale histology studies. This
places us into the
computational anatomy (CA) orbit problem for which constraints are inherited from an atlas that is diffeomorphic but not geometrically identical.
With the availability of dense brain atlases at many resolution scales \cite{Mai-Paxinos,Mori-Atlas,chuang2011mri,Mori-Miller-Atlas-Based-2013},
 methods to map atlas labels onto target coordinate systems are being ubiquitously deployed across neuroscience applications.
Since Christensen's early work
\cite{Christensen1996}, diffeomorphic transformation has become the de-facto standard as diffeomorphisms generate one-to-one and onto correspondences between coordinate systems.
Herein we focus on the diffeomorphometry orbit model \cite{Miller2014} of computational anatomy \cite{GrenanderMiller1998}, where the space of dense
volume imagery is modelled as a Riemannian orbit of an atlas under the diffeomorphism group.
We use the large deformation diffeomorphic metric mapping (LDDMM) algorithm first derived for dense imagery by Beg 
\cite{Beg2005} to retrieve the unknown high-dimensional reparameterization of the template coordinates.

Of course,
for the histological stacking problem solved here, the interesting twist is the augmentation of the random orbit model with 3 rigid motion dimensions for each target section.
At 20 $\mu$m, this implies as many as 500 sections augmenting the high-dimensionality of the diffeomorphism space to include as many as 1500 extra dimensions for planar rigid motions
for restacking. Here lies the crux of the challenge.
To accommodate the high-dimensionality of the unknown rigid motions, the space of stacked targets is modelled to have finite-squared energy Sobolev norm, which enters the problem as a prior distribution restricting the roughness of the allowed
restacked volumes.
 The variational method jointly optimizes over the high-dimensional diffeomorphism associated to the atlas
reparameterization and the high-dimensional concatenation of rigid motions associated to the target.
\section{METHODS}
\subsection{The Log-Likelihood Model of the Histology Sectioning Problem}
\begin{figure}[H]
\vspace{-0.5cm}
\hspace{1.8cm}
\includegraphics[width=0.7\textwidth]{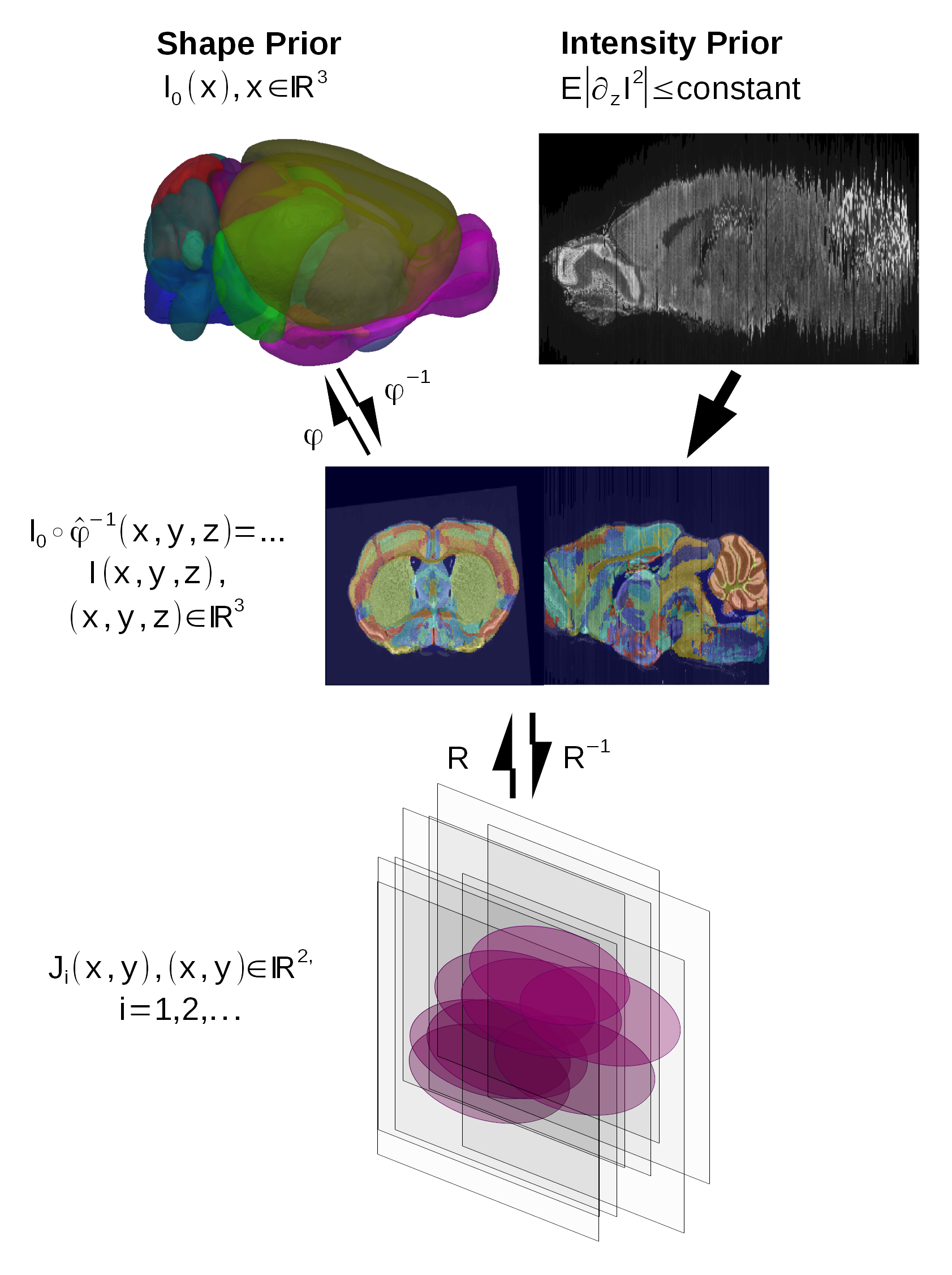} \\
\vspace{-1cm}
\caption{The histological sectioning model; the template $I_0$, the mouse brain in the orbit $I\in \mathcal I$ and observed histological sections  $J_i,i=1,\dots,n$. 
The Sobolev image intensity prior and the shape prior are depicted in the top row. 
 The model shows the template and mouse brain as elements of the same orbit $I_0,I \in \mathcal I$,  so there exists diffeomorphism $I=I_0 \circ \varphi^{-1}, \varphi \in \textit{Diff}$.  }
\label{fig:pipelineFigure}
\vspace{-0.5cm}
\end{figure}

Figure \ref{fig:pipelineFigure} shows the components of the model for the histology stacking problem. 
We define the mouse brain to be sectioned as a dense three-dimensional (3D) object $I(x,y,z), (x,y,z) \in \R^3$, modelled to be a smooth deformation of a known, given template 
$I_0  $ so that $I= I_0 \circ \varphi^{-1}$ for some invertible diffeomorphic transformation $\varphi$.
The Allen Institute's mouse brain atlas \cite{allen-mouse-brain-atlas-2007} (CCF 2017) is taken as the template.
Distinct from volumetric imaging such as MRI which delivers a dense 3D metric of the brain, the histology procedure  (bottom row, Figure \ref{fig:pipelineFigure}) consisting of sectioning, staining, and imaging 
generates a jitter process which randomly translates and rotates the stack sections. Denote the rigid motions  acting on the 2D sectioning planes $R_i: {\mathbb R}^2 \rightarrow {\mathbb R}^2$,
\begin{equation}  
R_i (x,y) =( \cos \theta_i  x+ \sin \theta_i  y +t_i^x, 
-\sin \theta_i x+ \cos \theta_i  y   + t_i^y) \ ,  \   (x,y) \in {\mathbb R}^2\ ,
\end{equation}
with $\theta_i$ the rotation angle and $(t_i^x, t_i^y) \in \mathbb{R}^2$ the translation vector in section $i$. 
The histology stack $J_i (x,y), (x,y) \in {\mathbb R}^2, i=1,\dots,n $, is a sequence of 2D jittered image sections
under smooth deformation of the atlas in noise:
\begin{equation}  J_i \circ R_i(x,y) = I_0 \circ \varphi^{-1}(x,y,z_i) + \textit{noise}(x,y),  \ \   (x,y) \in {\mathbb R}^2 \ . 
\label{basic-model-equation}
\end{equation}

Modeling the photographic noise as Gaussian and conditioning on the sequences of jitters $R_i,i=1,\dots,n$ and atlas deformation $I=I_0 \circ \varphi^{-1},\varphi \in \textit{Diff}$,  the photographic sections $J_i$ are a sequence of conditionally Gaussian random fields
with log-likelihood
\begin{eqnarray}
\ell(v,R;J) &=& \sum_i  
\left( - \alpha_i
\int_{{\mathbb R}^2} | J_i\circ R_i(x,y)  -  I_0 \circ \varphi^{v,-1} (x,y,z_i) |^2 dx dy \right) .
\label{log-likelihood-J-R}
\end{eqnarray}
Here $\alpha_i$ is a weighting factor dependent on the noise of each section such that damaged sections can be weighted; $v$ denotes the vector field which indexes the deformation as a diffeomorphic flow (see below).
\subsection{The Priors: Diffeomorphisms and Sobolev Smoothness of Images}
The parameterization of the histology pipeline augments the standard random orbit model of computational anatomy with the rigid-motion dimensions of the random jitter sectioning process.
The unknowns to be estimated become
$(R_1,\dots,R_n, \varphi) \in {\mathbb R}^{3n} \times \textit{Diff}$ for $n-$sections.
At $20 \mu$m then $n=500$ implying the nuisance rigid motions are of
high dimension $O(1500)$. 
The solution space must be constrained. We use priors on the deformations and on the rigid motion stacking of the images.
\\\\
{\bf The Diffeomorphism Prior:}
The histological stacking
constrains the brains as smooth transformations of the template, where the diffeomorphisms are generated as diffeomorphic flows $\varphi_t \in \textit{Diff}$ \cite{GrenanderMiller1998},
 solving the ordinary differential equation
\begin{equation}
\dot \varphi_t = v_t \circ \varphi_t , t \in [0,1], \ \varphi_0 = \textit
{identity} \ ,
\label{Lagrangian-equation}
\end{equation}
with $v_t$ the Eulerian velocity taking values in $\mathbb{R}^3 $, $\textit{identity}$ the identity mapping.
The top row of Figure \ref{fig:pipelineFigure}a shows that each $\varphi$ has an inverse and that the random orbit model assumes any individual brain $I \in \mathcal I$ can be generated from the exemplar under the action of the diffeomorphism, so that for some $\varphi \in \textit{Diff}$, $I= I \circ \varphi^{-1} $.
To score the distances between mouse brain coordinate systems and reject outlier solutions
we use geodesic  flows minimizing metric length \cite{Miller2006}.
  Large deviations as measured by the diffeomorphometry metric \cite{Miller2014} from template atlas to target mouse brain are thus removed from the solution space. 
The vector fields are modeled to be in a reproducing kernel Hilbert space (RKHS) 
$(V,\|\cdot\|_V)$, supporting one continuous spatial derivative, and having geodesic length between coordinate systems determined by the norm-square $\| v \|_V^2$ of the RKHS:
\begin{equation} \| v \|_V^2 = \sum_{i=1}^3 \int_{\R^3} ((-\nabla^2+1)^2 v_i(x,y,z))^2 dxdydz < \infty \ .
\label{RKHS-equation}
\end{equation}
This square-metric is used as a quadratic potential for the smoothness prior between images $I,I ^ \prime \in \mathcal I$  \cite{Miller-Younes-2001,Miller2015} minimizes the action
\begin{equation}
\rho^2 (I,I^\prime) = \min_{\varphi: \varphi_0 = \id, \varphi_1 \cdot I=I^\prime} \int_0^1 \| v_t \|_V^2 dt
\ .
\label{metric-distance}
\end{equation}
See Appendix \ref{geodesic-appendix-section} for the explicit equations for geodesics satisfying the Euler-Lagrange equations \cite{Miller2002,Miller2006} and Appendix \ref{Green-kernel-rkhs} for the matrix Green's kernel.

We use the notation $\varphi^{v}$ to emphasize the dependence of the diffeomorphism and the geodesic metric on the vector field $v$.
Strictly speaking, the group generated by integrating \eqref{Lagrangian-equation} with
finite norm $\| \cdot \|_V$ is both dependent on the norm of $V$ as well
as a subgroup of all diffeomorphisms; we shall suppress that technical detail in the notation.
\\\\
{\bf The Prior Distribution on Image Smoothness:}
To score the maximum a-posteriori (MAP) reconstruction of the rigid motions acting on the stack, we exploit a smoothness prior on the reconstructed histology stack
which enforces the fact that anatomical structures are smooth and continuous.
We model the images as arising from a smooth ``Sobolev'' or RKHS $I \in H^k$ supporting derivatives $ \partial_h f= \frac{\partial^{h_1+h_2+h_3}}{ \partial x^{h_1}  \partial y^{h_2} \partial z^{h_3}} f $  that are square integrable,
 with norm:
\begin{equation}
\| I \|_{H^k}^2 =
\sum_{h_1,h_2,h_3:|\sum_{i=1}^3 h_i| \leq k}
\int_{\R^3} |\partial_h I (x,y,z) |^2 dxdydz
 \ .
\label{Hk-quadratic-form}
\end{equation}

This is a quadratic form for a Gaussian random field prior on the dense histology stack with zero mean and covariance dependent on the squared norm $\| I \|_{H^k}^2$.
For the purpose of stacking, the z-axis sections are sparse $20-40 \mu$m; the differential operators $\partial_h$ are implemented via the
difference operator along the sectioning z-axis  (see Eqn. \eqref{discrete-difference-oeprator}).
The Gaussian field has covariance determined by the difference operators; see 
\cite{lanterman-miller:bayesian-segmentation-2000} for example.
We define the mixed differential-difference operator $D_h$ as the centered difference for
the z-partial derivatives, 
\begin{equation}
D_h f (x,y,z)= \partial_{h_1,h_2} \left( \frac{ f(x,y,z+\Delta/2) - f(x,y,z-\Delta/2)}{\Delta} \right) \ . 
\label{discrete-difference-oeprator}
\end{equation}
The gradient is forced to 0 at the boundaries of the image.

\subsection{MAP, Penalized-Likelihood Reconstruction}
Model the random sectioning with section-independent jitter as a product density $\pi(R) = \prod_{i} \pi(\theta_i,t_i^x,t_i^y) $, the priors centered at identity.
Generating MAP estimates of the rigid motions generates the MAP estimator of the histology restacking problem denoted as
$$ I^R (x,y,z_i) = J_i \circ R_i(x,y) , (x,y) \in {\mathbb R}^2,  \ \ i=1,\dots,n \ . $$
Since the diffeomorphisms are infinite dimensional, the maximization of the log-likelihood function with respect to a function with the deformation penalty is termed the "penalized-likelihood estimator".
Conditioned on the known atlas, the augmented random variables to be estimated are $(R_1, \dots, R_n, \varphi ) \in ({\mathbb R}^{3n} \times \textit{Diff}) $.

\begin{problem} [\textbf{MAP, Penalized-Likelihood Estimator}]\ \\
Given histology stack $J_i (x,y),(x,y)\in {\mathbb R}^2, i=1,\dots \ $ and reconstructed
stack $I^R(\cdot,z_i) = J_i \circ R_i(\cdot), i=1,\dots,n$ modelled as conditionally Gaussian random fields conditioned on jitter and smooth dormation of the template.
The joint MAP, Penalized-Likelihood  estimators $\arg \max_{R,v} \log \pi(R,v|J)$ given by 
\begin{eqnarray}
\label{MAP-energy}
 \argmax_{R,v} &&
-\frac{1}{2} \int_0^1 \|v_t \|_V ^2 dt 
-\frac{1}{2}\sum_i \| D_{h} I^R (\cdot,z_i) \|_2^2
\\
&& +
 \sum_i  
\left( \log \pi(R_i) - \alpha_i  
\| I^R(\cdot,z_i)  -  I_0 \circ \varphi^{v, -1}(\cdot,z_i) \|_2^2 \right)  \  .
\nonumber
%\int_{X_i} | J_i \circ R_i (x,y) -  I_0 \circ \varphi^{v -1}(x,y,z_i)|^2 dx dy \  ,
\end{eqnarray}
The  MAP, Penalized-Likelihood estimators
satisfy
\begin{equation}
\begin{cases}
&  R^* = 
\argmax_{R_i, i=1,\dots} \sum_i \left( \log \pi(R_i) -\frac{1}{2}
 \| D_{h} I^R (\cdot,z_i) \|_2^2 \right.
\\
&\ \ \ \ \ \  \ \ \ \ \ \ \ \ \ \ \ \ \ \ \ \ \ \ \ \ \ \ \ \ \ \ \ \ \ \ \  \left.
- \alpha_i \| I^R(\cdot,z_i) -  I_0 \circ \varphi^{ v^*, -1}(\cdot,z_i)\|_2^2  \right)  \ ,
\ \ \ \ 
\nonumber
\\
&
\begin{aligned}[b]
v^* = \argmax_{v}-\frac{1}{2} \int_0^1 \|v_t \|_V ^2 dt -
\sum_i  
\alpha_i
\| I^{R^*}(\cdot,z_i)  -  I_0 \circ \varphi^{v, -1}(\cdot,z_i) \|_2^2
\end{aligned}
\end{cases}
\end{equation}
with $\| \cdot \|^2_2$ denoting the norm per z-axis section:
\begin{equation} \| f (\cdot,z_i) \|_2^2 = \int_{\mathbb{R}^2} f(x,y,z_i)^2 dxdy \ . 
\label{square-integral-xyplane}
\end{equation}
\end{problem}
We call this the \textbf{atlas-informed} model. The first two prior terms of \eqref{MAP-energy} control the smoothness of template deformation and the realigned target image stack, with
the third keeping the rigid motions close to the identity.
The last term is the ``log-likelihood'' conditioned on the other variables.

The optimization for the $R^*$ rigid-motions is not decoupled across sections
because of the smooth diffeomorphism of the LDDMM update and the Sobolev metric represented through the difference operator across the $z-$ sections.
The optimization of the vector field $v^*$
corresponds to the LDDMM solution of Beg \cite{Beg2005}.

The principal algorithm used for solving this joint MAP-penalized likelihood problem alternates between fixing the rigid motions and solving LDDMM and fixing the diffeomorphism and solving for the rigid motions. This is described in the Methods section.

When there is no atlas available this is equivalent to setting $\alpha_i$ small and becomes a MAP rigid motion restacking of the sections:
%The MAP estimator becomes
$$ 
 \argmax_{R_i, i=1,\dots } \sum_i \left(\log \pi(R_i)  -\frac{1}{2}
 \| D_{h} I^R (\cdot,z_i) \|_2^2 \right) \ .
$$
We term this the \textbf{atlas-free} model. The gradient of the rigid motions with respect to the components of translations $t^x,t^y$ and rotation $\theta$ is defined in Appendix
\ref{Gradients-atlas-free-model-appendix}. 
The registration is not independent across sections due to coupling through the Sobolev metric.

\subsection{Iterative Algorithm for Joint Penalized Likelihood and MAP Estimator}
Here we describe the details of the algorithm used for solving for the MAP/penalized--likelihood problem of section 2.3. The algorithm alternately fixes the set of rigid motions while updating LDDMM and fixes the diffeomorphism while updating the rigid motions.
\begin{implementation}
\label{atlas-informed-algorithm}
\text{}
\begin{algorithmic}
\STATE 0. Initialize $\varphi^{new},R^{new} \leftarrow \varphi^{init},R^{init}$, $ I^{old} \leftarrow J \circ R^{init}$:
\STATE 1. Update $\varphi^{old} \leftarrow \varphi^{new},R_i^{old} \leftarrow R^{new}_i$, $I^{old}(\cdot,z_i) \leftarrow I^{new}(\cdot,z_i), i=1,\dots $.
\STATE 2. Update LDDMM for diffeomorphic transformation of atlas coordinates:
\STATE \begin{eqnarray} v^{new} &=& \argmax_{v} \ 
% \dot \varphi_t 
%= v_t \circ \varphi_t 
-\frac{1}{2}
\int_0^1 \| v_t \|_V^2 dt -\sum_i \alpha_i
\| I^{R-old}(\cdot,z_i)
-  I_0 \circ \varphi_1^{v-1}(\cdot,z_i) \|^2  \  ,
\ \ \ \ \ \ \
\label{LDDMM-minimization}
\\
\varphi^{new} &=& \int_0^1 v_t^{new} \circ \phi_t^{new} dt + \id 
\  .
\nonumber
\end{eqnarray}
%Here $K$ is a Gaussian kernel or Green's kernel of $V$
\STATE 3. Deform atlas $ I_0 \circ \varphi^{new-1}$ and generate new histology image stack:
\STATE \begin{eqnarray}
& R^{new}&=
\label{rigid-motion}
\arg \max_{R_i,i=1,\dots}
% \sum_i \left( \log \pi(R_i)-\frac{1}{2}
 \sum_i \left( \log \pi(R_i) \right. 
\\ && \ \ \ \ \ \ \ \ \ \ \left.
-\frac{1}{2}
 \| D_{h} I^R(\cdot,z_i) \|_2^2 - \alpha_i \| I^R(\cdot,z_i) -  I_0 \circ \varphi^{new -1}(\cdot,z_i)\|_2^2  \right)\ ; \ \ \ \ 
\nonumber
\\
& I^{R-new}(\cdot,z_i) & =  J_i \circ R_i^{new} (\cdot)  \ , i =1 \dots
\nonumber
\end{eqnarray}

\STATE 4. Return to Step 1 until convergence criterion met.

\end{algorithmic}
\end{implementation}
The form of the gradients for the rigid motions is given in Appendix
\ref{Gradients-random-orbit-appendix}.
The LDDMM update solutions are given by Beg \cite{Beg2005}.

\subsection{Software Implementation}
The algorithm described above is applied to Nissl histological stacks using the Allen Institute's mouse brain atlas as a template. The Allen Mouse Brain Atlas is a micron-scale atlas that includes annotated Nissl-stained images at 10, 25, 50, and 100 $\mu$m voxel resolution, with 738 labeled compartments in the annotation.  

Atlas mapping is computed on the Nissl-stained histological image stack showing the clear definition of anatomical boundaries. 
The associated fluorescent tracer images are transformed to the Nissl stack so that the atlas subvolume labels can be cast onto the new modality. The fluorescent and Nissl images are registered within animals by applying rigid registration based on a mutual information cost function.

\begin{figure}[H]
\vspace{-2cm}
\hspace{-2cm}
\includegraphics[width=1.25\textwidth]{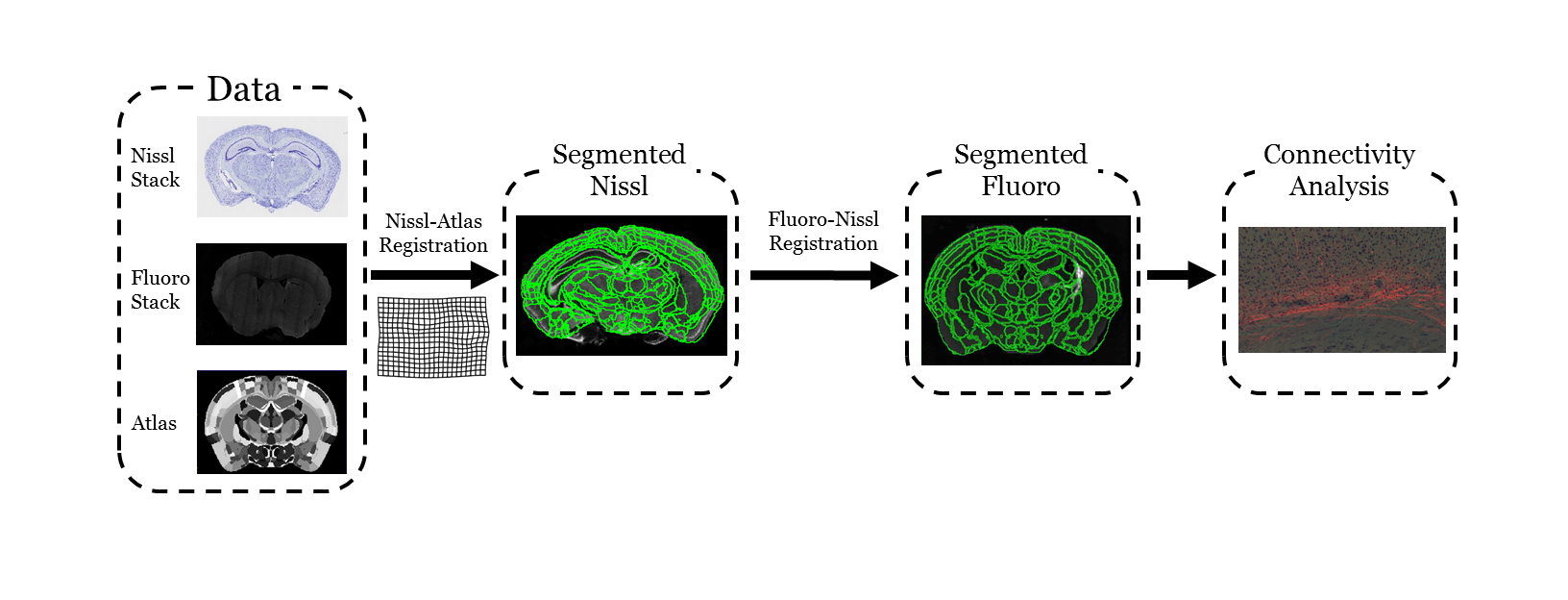}
\vspace{-1.5cm}
\caption{Histology registration pipeline workflow}
\label{fig:softwarepipelineFigure}
\end{figure}

A software pipeline which performs start-to-finish registration operations was implemented on a high performance computing cluster for atlas-mapping and histology restacking on the Mouse Brain Architecture data. To date, the pipeline has been successfully run on over 1000 MBAP brains. The general pipeline workflow is illustrated in Figure \ref{fig:softwarepipelineFigure}.
The pipeline begins with a pre-selection of the atlas model. After affine registration of the Nissl sections is applied, the algorithm of section 2.4 is applied to compute the mapping between atlas and target coordinate spaces.
\noindent In our application, we apply a two channel LDDMM \cite{CeritogluMultiModal2009}
algorithm
for the optimization with respect to $\varphi$, where the first channel is the Nissl-stained grayscale image, and the second channel is a mask of the brain tissue with ventricles and background set to a pixel value of zero. The brain mask for each brain stack is automatically generated by thresholding at an estimated background intensity value and applying morphological opening and closing for denoising. The threshold value is estimated by a RANSAC-like procedure over the image histogram, assuming a normal distribution of intensity values in the image foreground. A first-order Sobolev-norm (see below) is used for the smoothness constraint regularization of the
histology stack. In order to accommodate for sections damaged by the histology process or structures excluded from imaging, the objective functions in all parts of the algorithm are optimized with respect to only the image data that exists. Essentially, this is a masking procedure on the cost function that allows matching between a whole atlas brain and some target which is a partial, or subset of a whole brain.

After registration of the structural Nissl image, the fluorescence volume is registered to its corresponding Nissl volume. The registration is restricted to rigid motions on each individual section. The optimization bears a similar form to equation \eqref{rigid-motion} with the squared error matching term replaced with mutual information in order to account for the different modalities of the template and target histology stack. Once fluoro-to-Nissl registration is complete, the Nissl segmentation can be applied to the fluorescence image.

\section{RESULTS AND DISCUSSION}
\subsection{Validation on Simulated Reconstructions}
\subsubsection{The Phantom with Curvature}
\begin{figure}[H]
\hspace{-1.5cm}
\includegraphics[width=1.2\textwidth]{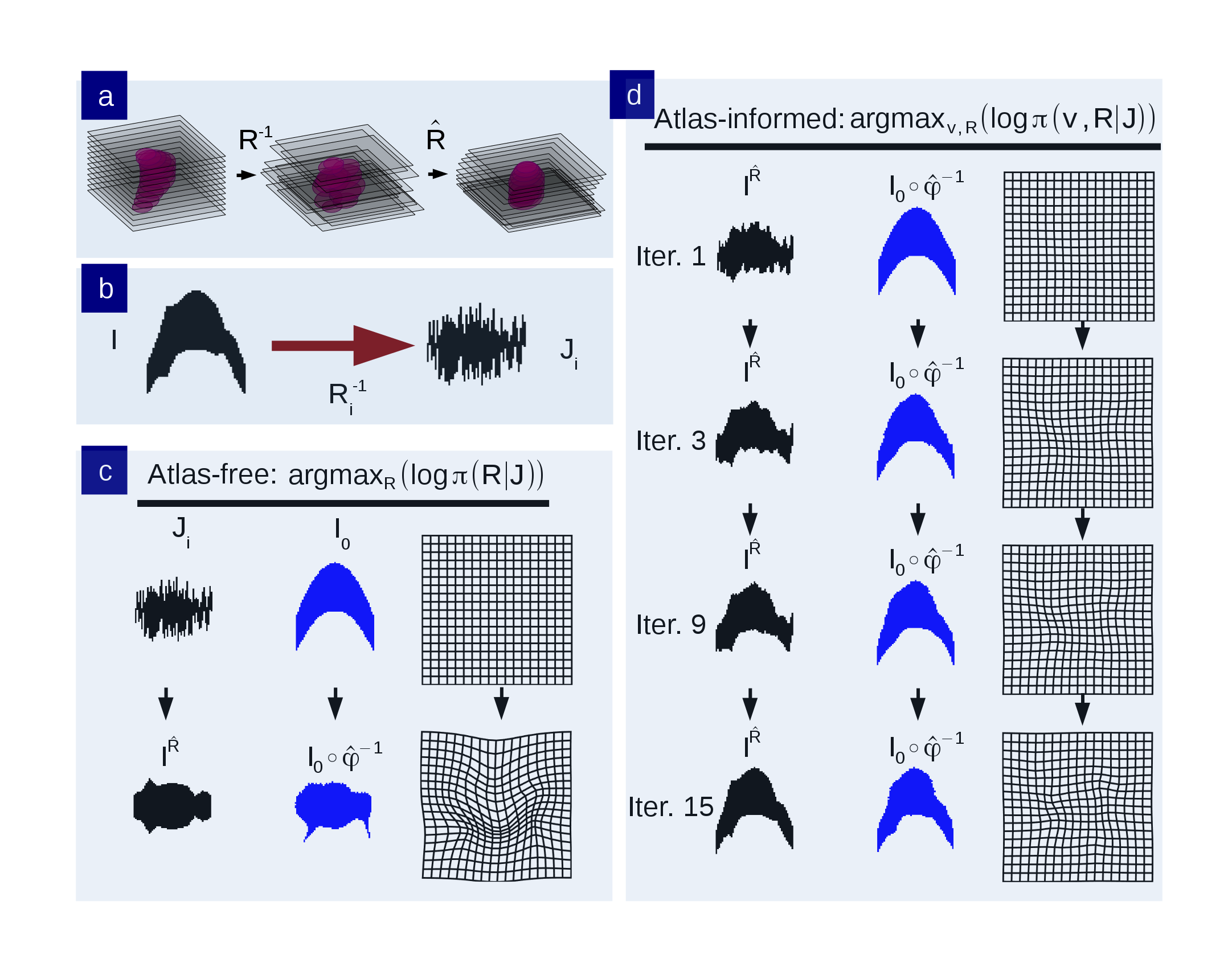}
\caption{a) An illustration of the classic curvature reconstruction problem. b) The unobserved 3D-phantom is randomly sectioned and observed as $J_i,i=1,\dots, n$. c) Reconstruction of the histological stack using the atlas-free method. The top row shows the histological stack and atlas. The bottom row shows the reconstructed histological stack $I^{\hat{R}}$ alongside the deformed phantom atlas $I=I_0 \circ \varphi^{-1}$ which has been mapped to histological sections, and the diffeomorphic change of coordinates $\hat \varphi^{-1}$. d) Reconstruction of phantom using the atlas-informed model. Each row depicts iterations of the reconstructed histological stack $I^{\hat{R}}$ alongside the deformed atlas $I=I_0 \circ \hat \varphi^{-1}$ and deformed coordinates.
The bottom row shows the convergence point of the algorithm.}
\label{fig:bananaMovieA}
\end{figure}

The model was applied to binary image phantoms in order to examine the ``curvature'' problem in which a 3D curved object cannot be accurately reconstructed after being sectioned. This is illustrated in Figure \ref{fig:bananaMovieA}.
We produced sections through the 3D phantom, applying the atlas-free and the atlas-informed models. 
%where neither model has any prior knowledge of the unobservable ground truth target $I$ (shown in Figure \ref{fig:bananaMovieA}b with its observation $J_i$).
The results from the atlas-free algorithm in which the sections are aligned based on the Sobolev smoothness followed by mapping of the atlas via LDDMM are summarized in Figure \ref{fig:bananaMovieA}c.
The atlas-free section alignment reconstructs the target stack,
demonstrating a cylindrical reconstruction rather than the curved template shape, followed by LDDMM alignment $I_0 \circ \varphi^{-1}$. 
This illustrates the curvature issue. The atlas coordinate grid is transformed significantly (bottom right of Figure \ref{fig:bananaMovieA}c) in order to match the target.
%Although the atlas has deformed significantly to match the target in the atlas-free model, 
Despite this significant deformation, there is some residual error in the atlas-to-target mapping with the remaining tendrils where the ends of the phantom did not shrink inwards. Here, the energy required to push the ends of the atlas inwards were greater than the potential image matching improvement.

Shown in Figure \ref{fig:bananaMovieA}d is the atlas-informed solution.
The bottom row of Figure \ref{fig:bananaMovieA}d 
shows that simultaneously solving for reconstruction and registration parameters allows for more consistent stack reconstruction of the target resulting from the influence of the smooth deformation of the template onto the target in the joint solution. 

\begin{figure}[H]
\hspace{-1.2cm}
\includegraphics[width=1.1\textwidth]{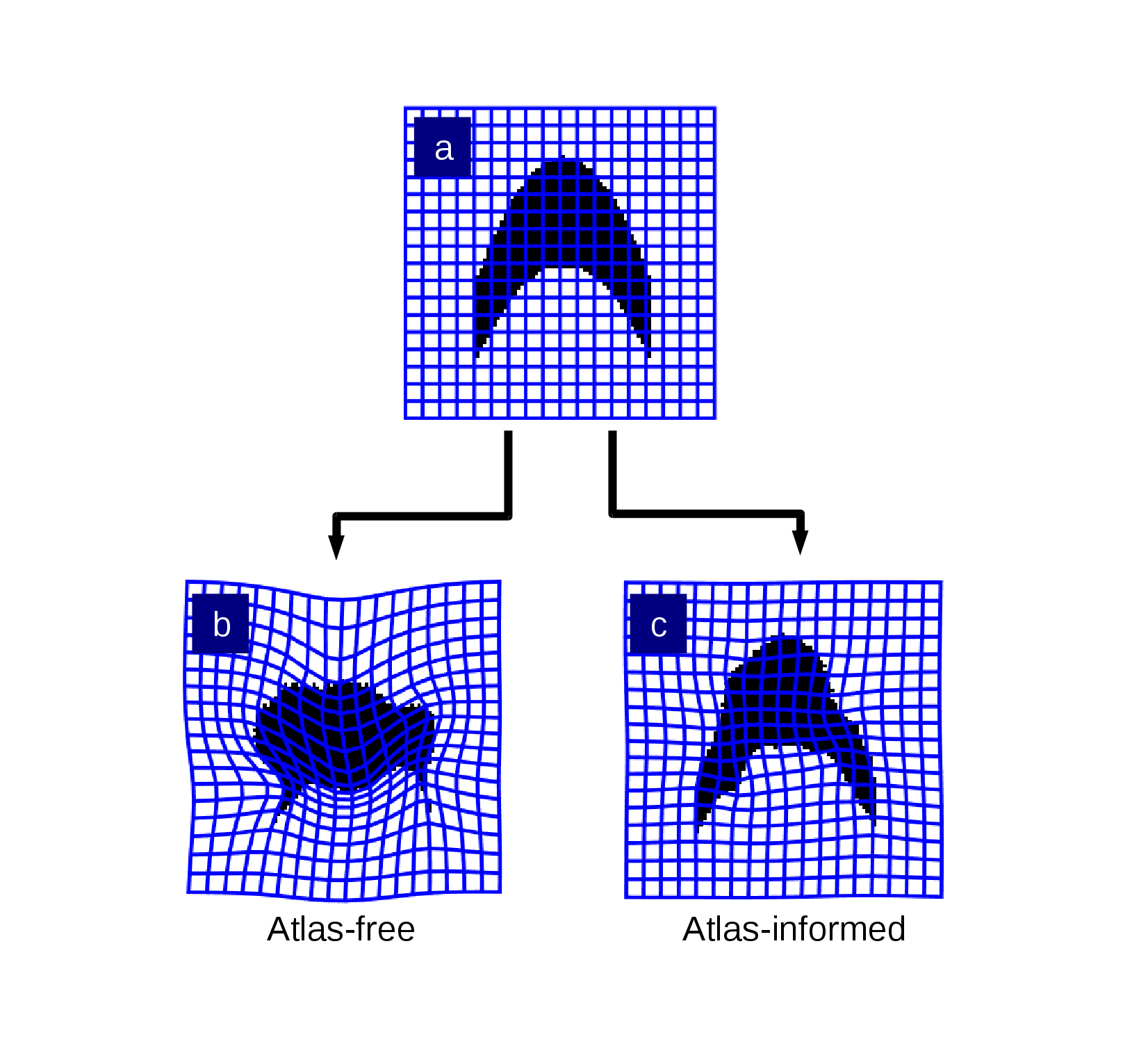}
\vspace{-1cm}
\caption{Transformed grids illustrating the difference in the mapping deformation from the atlas-free methods from (A) to histology stack target (B) versus the atlas-informed algorithm which produces (C).}
\label{fig:bananaFigDeformedGrids}
\end{figure}

These results are depicted by the difference in the motions of the atlas coordinate grids when deforming onto the targets in Figure \ref{fig:bananaFigDeformedGrids}. Tandem optimization of section alignment parameters and diffeomorphisms produces a nonlinear mapping with lower metric cost (Fig \ref{fig:bananaFigDeformedGrids}c is less warped than Fig \ref{fig:bananaFigDeformedGrids}b).
\subsubsection{Jittering the Allen Atlas}
\begin{figure}[H]
\vspace{-1.5cm}
\centering
\includegraphics[width=0.9\textwidth]{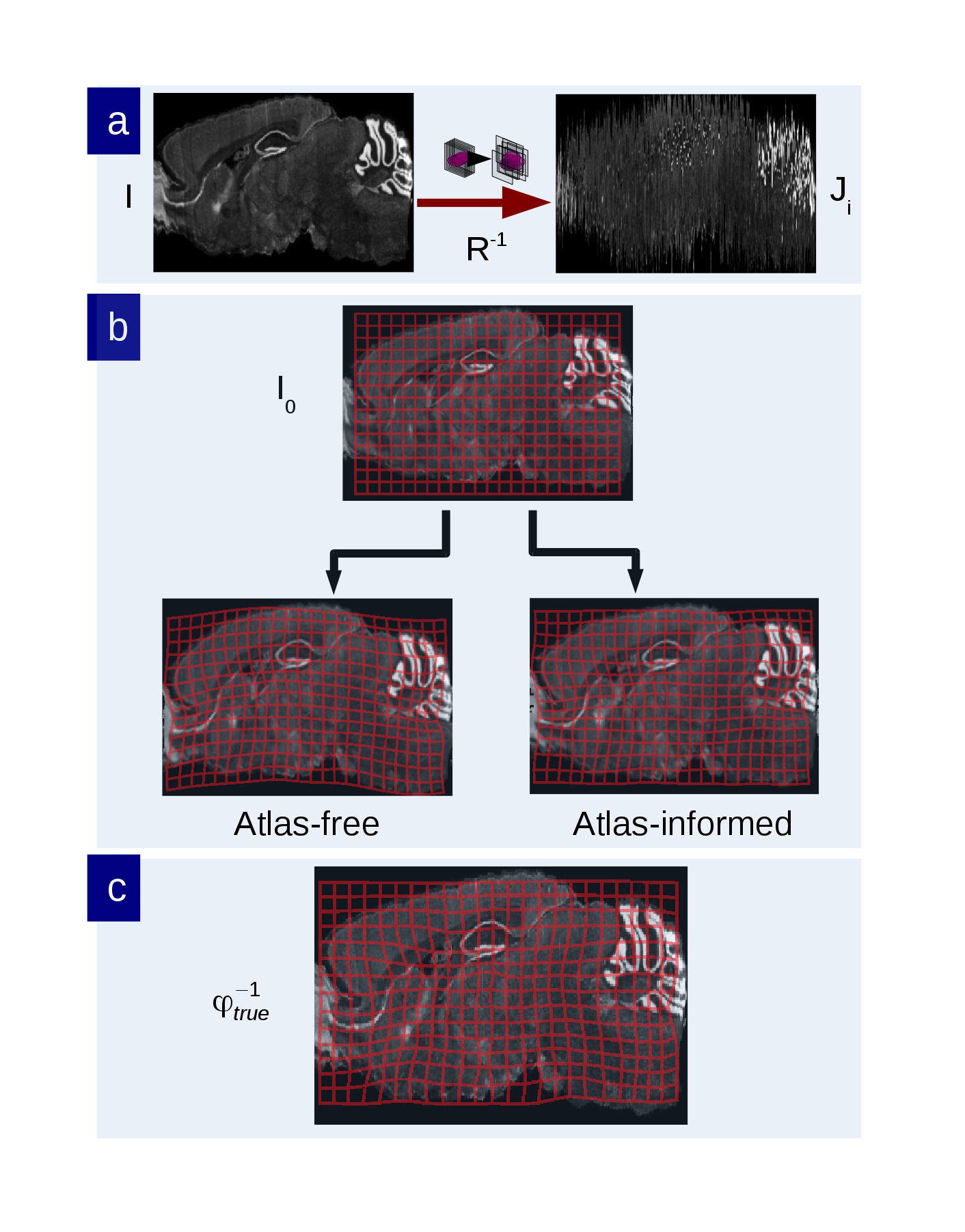}
\vspace{-1cm}
\caption{Atlas phantom simulation to validate recovery of sectioning parameters and diffeomorphic shape difference. a) The ground truth target $I$ is sectioned to generate the observed target $J_i$.
b) Transformed grids illustrating the brain phantom atlas (top) shown mapped onto the histological stack using the atlas-free algorithm (bottom left) and the atlas-informed algorithm (bottom right). c) The ground truth diffeomorphism to be recovered.}
\label{fig:atlasFig}
\end{figure}

A similar experiment was performed using the Allen mouse brain atlas as the 3D phantom. A target histology stack was generated by sectioning the Allen atlas in simulation and applying random rigid transforms to its coronal sections. The atlas images were sampled at 40 $\mu$m isotropic voxels.
This is depicted in Figure \ref{fig:atlasFig}a. A simulated atlas was generated by applying a given random diffeomorphism to the Allen atlas. This random diffeomorphism is depicted in Figure \ref{fig:atlasFig}c.
The histology stacks were then reconstructed and diffeomorphic transformations generated between the atlas and target stacks using both models, intending to recover both the unknown rigid transforms from Figure \ref{fig:atlasFig}a and the unknown diffeomorphism from Figure \ref{fig:atlasFig}c.
Figure \ref{fig:atlasFig}b shows the atlas-free method method (bottom left) compared
to the atlas-informed method (bottom right).
The atlas-informed method nearly reproduces the original coordinates whereas the atlas-free method drifts away from the original coordinates. Note that although the diffeomorphisms are not identical, this does not necessarily indicate segmentation error as small differences in stack alignment can be compensated for by nonlinear registration during atlas-mapping.

\subsubsection{Simulated Bias and Variance Statistics}
\begin{figure}[H]
\centering
\includegraphics[width=0.6\textwidth]{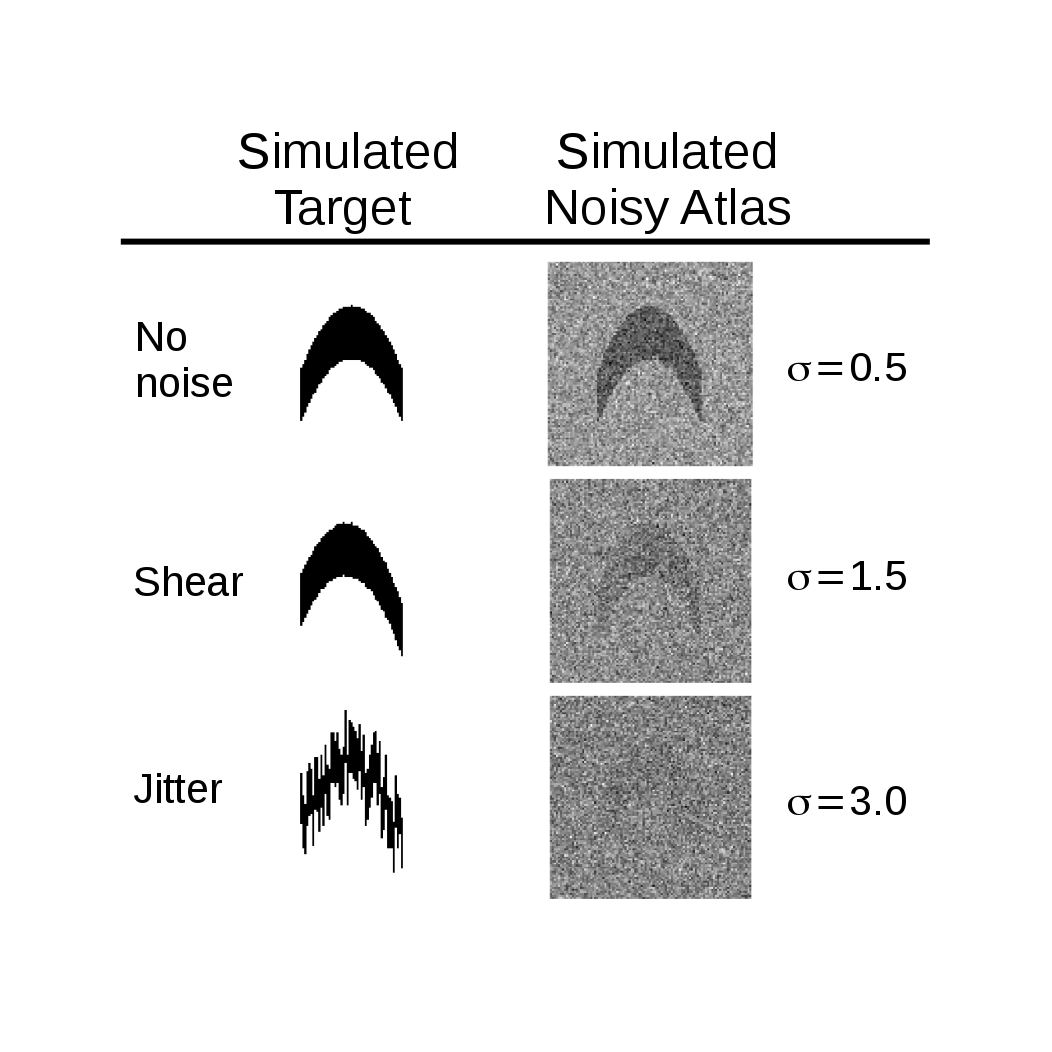}
%\vspace{-1.25cm}
\caption{Left column shows phantom for identity, shearing, and jitter of
sections (successive rows); right column shows Gaussian white noise added to the atlas at various standard deviations. 
The jitter random rigid motions were normally distributed $(t_x,t_y) \sim \mathcal{N}(\mu = 0, \sigma^2 = 36 ), \theta \sim \mathcal{N}(\mu = 0, \sigma^2 = 100 )$ in pixel units. }
\label{fig:banana_figure_noise}
\end{figure}

\begin{figure}[H]
\vspace{-0.3cm}
\hspace{-0.2cm}
\includegraphics[width=1\textwidth]{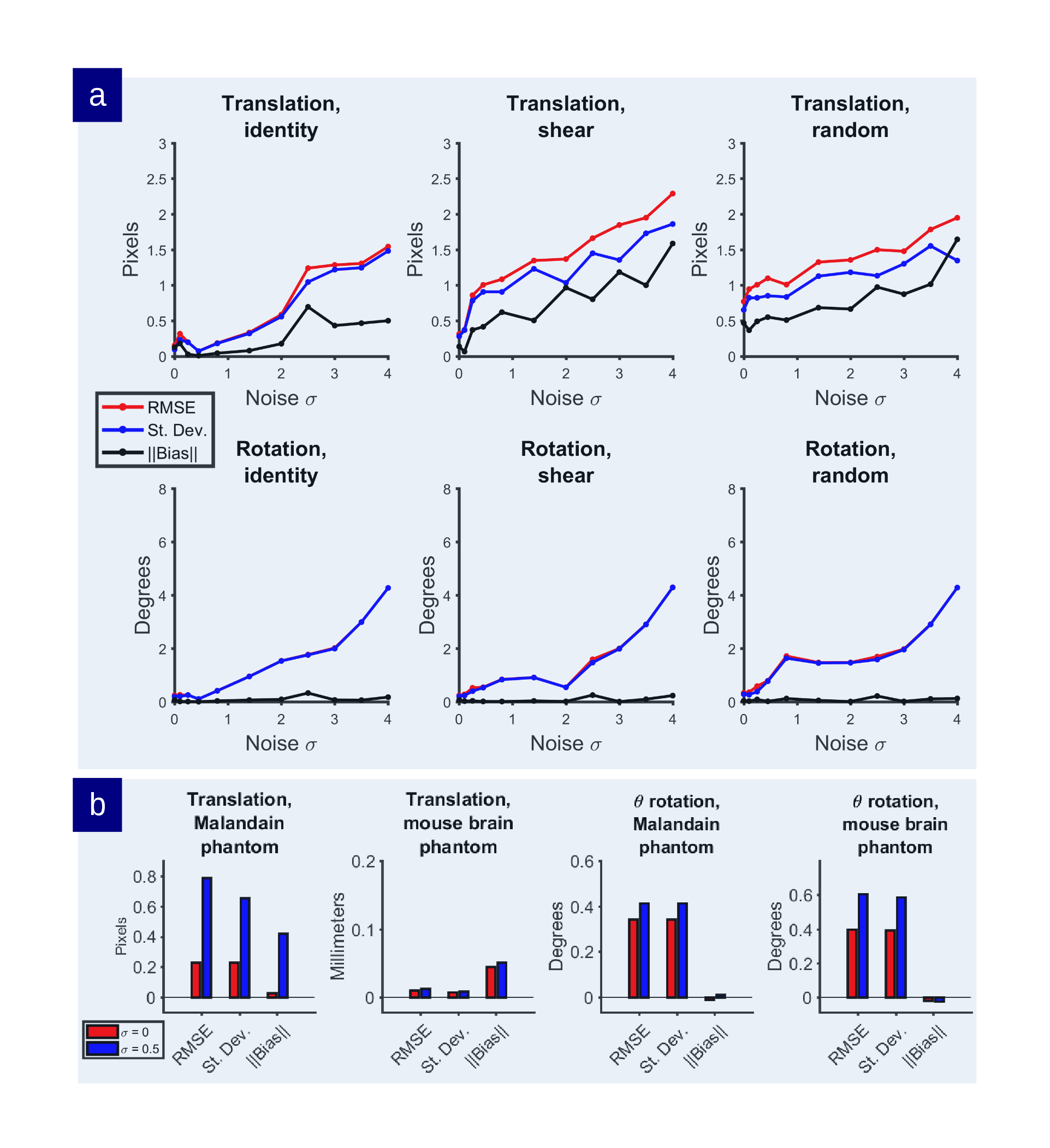}
%\vspace{-1.25cm}
\caption{a) Statistics on the translation-rotation estimators for noise levels varying initial conditions. b) Statistics on the rigid motion estimators where the section jitter was added in a random fashion.}
\label{fig:banana_stats_graph}
\end{figure}

Figure \ref{fig:banana_figure_noise} and \ref{fig:banana_stats_graph} show results quantifying the bias and viarance of the joint estimation of the diffeomorphism transformation and the rigid motion jitter in simulation. 
Eqn. \eqref{basic-model-equation} was simulated over a range of Gaussian white noise selections while simultaneously varying the jitter rigid motions of the sections along with multiple deformations of shearing applied to the template $I_0 $. 
Shearing produced images where each section was successively offset by 0.25 pixels in both x and y directions, cumulatively producing the ``shear'' effect illustrated in Figure \ref{fig:banana_figure_noise}. 
Figure \ref{fig:banana_stats_graph}b keeps the stack jitter fixed and varies the noise levels; Figure \ref{fig:banana_stats_graph}c varies the stack jitter. 
The RMSE, bias, and standard deviation of the estimated parameters were computed in each experiment and plotted as a function of error units versus noise level. 500 simulations per experiment were performed.
In each experiment, estimator accuracy is preserved up to high noise levels. At typical noise levels ($\sigma \leq 0.5$), we observe subpixel RMSE and small bias. Figure \ref{fig:banana_stats_graph}b shows that the rotation estimator is virtually  unbiased whereas the translation estimator does have small subvoxel bias. 
It is likely that more rotational error is accounted for by section realignment than deformable mapping, whereas both play a relatively balanced role in translation correction. Small motions are ill-posed in that small rigid-motions can accommodate small atlas deformation. 
Figure \ref{fig:banana_stats_graph}c (top row)
shows the case where the target stack are jittered. Estimator statistics are computed in each of these cases showing similar subpixel errors.

Simulations examining bias and variance were also run on the Allen atlas brain
as the phantom.  
The reconstruction RMSE observed in the brain phantom simulation (bottom row of Figure \ref{fig:banana_stats_graph}c) is lower than that observed in the simple curved phantom in pixels. It is likely that this is due to the presence of more contour lines in grayscale images versus binary images. These additional features allow for more accurate distinction of matching error than simpler images with small numbers of distinct level lnes. This is consistent with the demonstration
in \cite{Miller2006} showing that the stabilizer of the group corresponding to vector fields tangent to the level lines of the image cannot be uniquely identified or retrieved via any mapping methods that look at color or contrast of the image as the identifying feature.

\subsection{Mouse Brain Architecture Project data}
\begin{figure}[H]
%\vspace*{-0.9in}
\vspace{-1.5cm}
\hspace{2cm}
\includegraphics[width=0.7\textwidth]{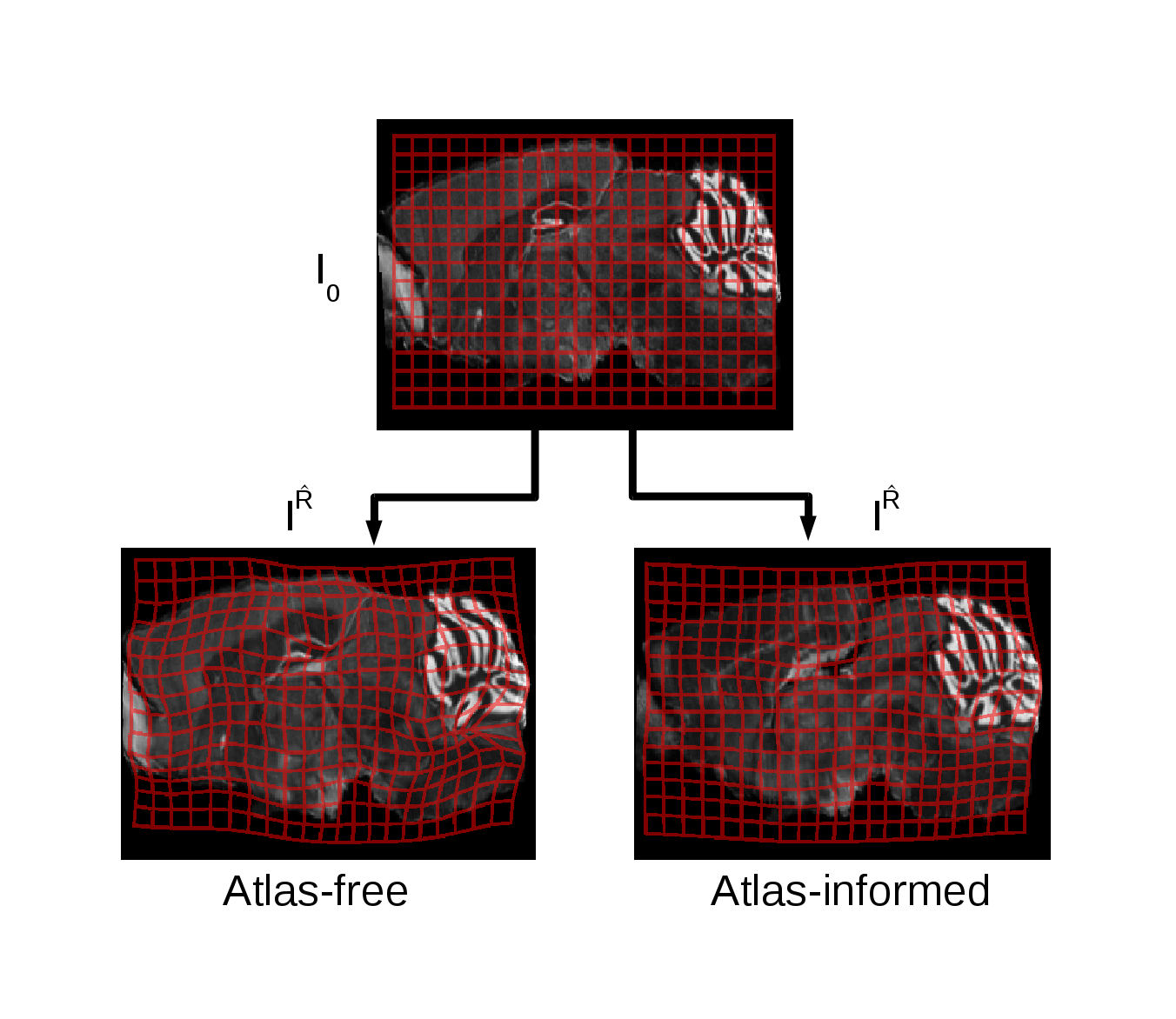}
\vspace{-0.5cm}
\caption{Transformed grids illustrating the difference in the mapping deformation from atlas (top) to target using the atlas-free method (bottom left) versus the atlas-informed method (bottom right), performed on real brain data from the MBA Project.}
\label{fig:targetFig_defgrids}
\end{figure}

A final experiment was conducted on brain data sampled from the MBAP database, using the Allen mouse brain as the atlas. We selected specific targets which were prone to poor registration results due to image intensity local minima. In particular, structures like the cerebellum tend to be difficult to register accurately due to their folded nature; one fold can easily be mistaken for the adjacent fold, and if the target and atlas are not well initialized, the deformation required to flow one fold onto another can have a high metric cost. We are also interested in inspecting lower-contrast structures like the corpus callossum, which may be poorly registered due to local minima in other nearby bright structures.
\begin{figure}[H]
%\vspace*{-0.9in}
\vspace{-1.5cm}
\hspace{2cm}
\includegraphics[width=0.7\textwidth]{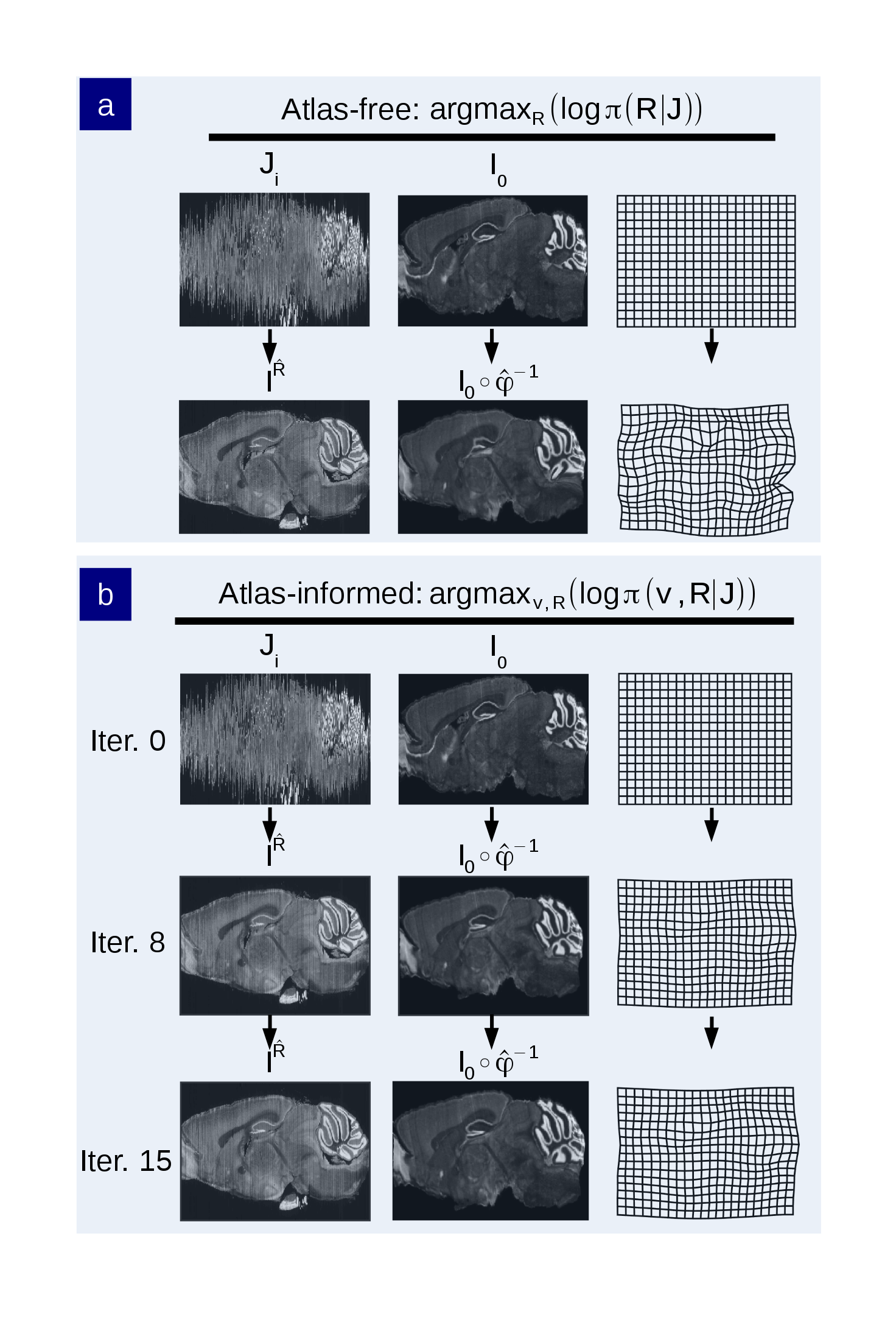}
\vspace{-0.5cm}
\caption{a) Reconstruction of an MBA Nissl-stained brain histological stack using the atlas-free method.
Top row shows the histological stack and Allen mouse brain atlas.
Bottom row shows the reconstructed histological stack $I^{\hat{R}}$ alongside the deformed phantom atlas $I$, and the diffeomorphic
change of coordinates $\hat \varphi^{-1}$.
b) Reconstruction using the atlas-free method.
Top row shows the histological stack and Allen mouse brain atlas.
Middle row depicts intermediate iterations of the reconstructed stack $I^{\hat{R}}$ alongside the deformed atlas $I_0 \circ \hat \varphi^{-1}$ and coordinate grid.
Bottom row shows the convergence point of algorithm. }
\label{fig:targetFig}
\end{figure}
The reconstructed histological target stack in the atlas-informed model shown in Figure \ref{fig:targetFig}a takes on the shape of the atlas but is prone to reconstruction artifacts.
The deformation grids produced by the atlas-informed mapping is much smoother and has many fewer wrinkles than the atlas-free mapping. This is seen clearly in Figure \ref{fig:targetFig_defgrids}.

\begin{figure}[H]
\vspace*{-0.7in}
\hspace*{0.15in}
\centering
\includegraphics[width=0.6\textwidth]{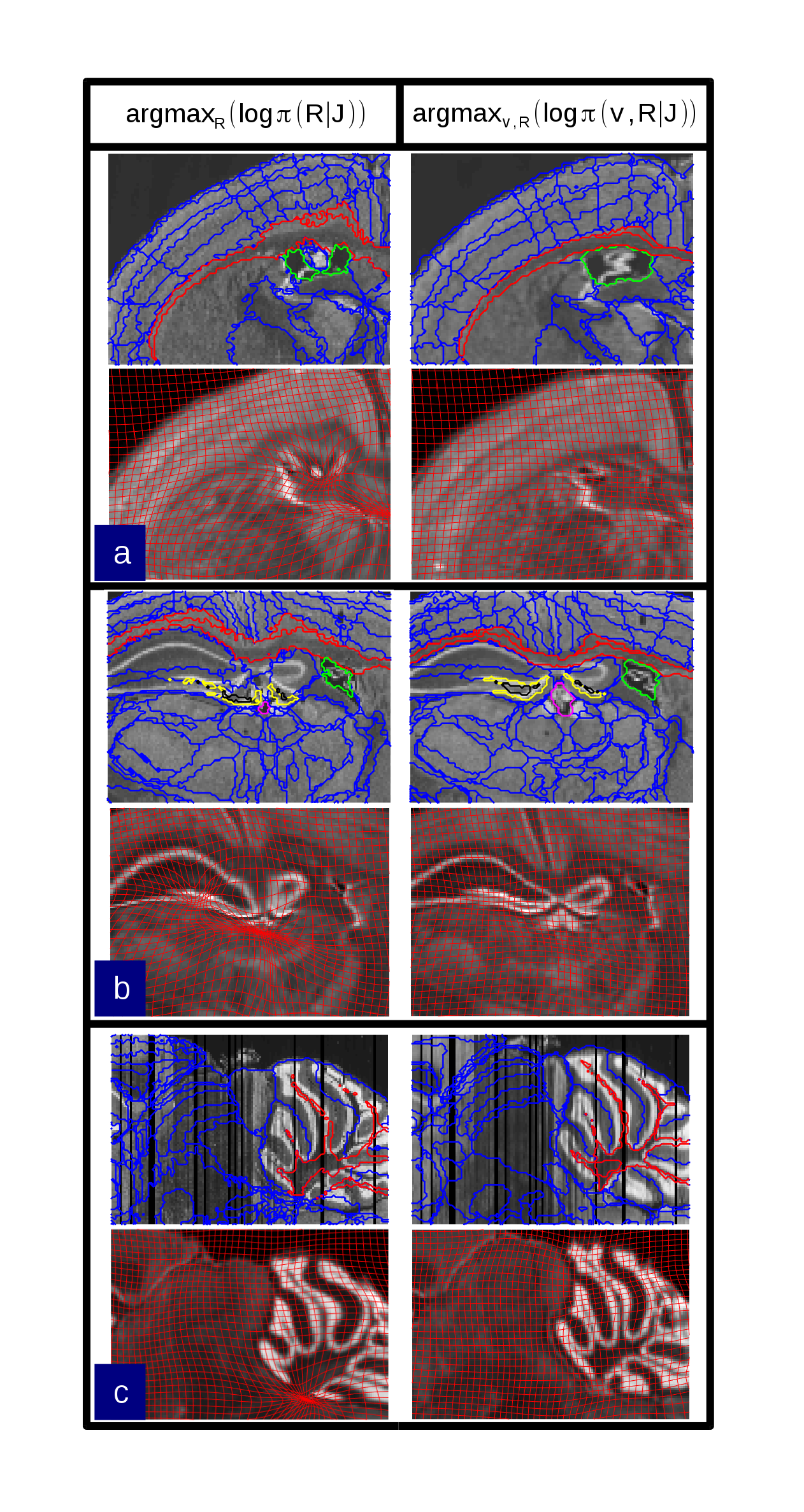}
\caption{Selected regions of the brain segmented by the atlas-informed and atlas-free models carry the label map from the Allen atlas. The left column shows several examples where optimization of the atlas-free solution is trapped in false minima due to folded or low-contrast structures. The right column shows correction by the atlas-informed algorithm. A) The corpus callossum and lateral ventricle. B) The dentate gyrus, corpus callossum, and lateral ventricle. C) The cerebellar white matter.}
\label{fig:segmentationFigure}
\end{figure}

Figure \ref{fig:segmentationFigure} shows examples of improved segmentations in selected regions of the brain.
The atlas-informed model generates more accurate segmentation results and produces smoother mappings as exhibited by the less wrinkled and distorted grids (bottom row b), showing more consistent results throughout the MBAP dataset.

\section{CONCLUSION}
This paper examines the CA random orbit model at the mesoscale for the stacking of sectioned whole brains coupled with mapping to annotated atlases.
The standard CA model has been expanded to include the O($3 \times \textit{n}$) extra rigid motion dimensions representing the planar histology sections.
%, assuming random orientation and translation jitter for each section.
The estimation procedure solved here simultaneously estimates the diffeomorphic change of coordinates between atlas and target histological stack, as well as the ``nuisance'' rigid motion parameters for each section in stack space.
This requires the introduction of a smoothness constraint on the target jitter simultaneous with LDDMM, which is enforced via a Sobolev metric, encouraging the reconstructed stack to be smooth by controlling the derivative along the cutting axis.

Results are shown demonstrating that the introduction of an atlas into the estimation scheme solves several of the classic problems associated with volume reconstruction, including the reconstructin of the curvature of extended structures. Since the atlas gives \textit{a priori} indication of the global shape, the tendency to remove distortions along the section axis is balanced against the desire to minimize the amount of deformation of the atlas onto the reconstruction.
The algorithm is shown to mediate this tension well.

%\section{METHODS}

\section*{Acknowledgments}
The authors would like to acknowledge Keerthi K Ram and Daniel Ferrante for all of their guidance in understanding the Mouse Brain Architecture Project datasets.
This work was supported by the G. Harold and Leila Y. Mathers Foundation, the Crick-Clay Fellowship, the H.N. Mahabala Chair, National Science Foundation Eager award 1450957, the Computational Anatomy Science Gateway as part of the Extreme Science and Engineering Discovery Environment (XSEDE) with grant number ASC140026, NIH DA036400, as well as the Kavli Neuroscience Discovery Institute supported by the Kavli Foundation.

\clearpage

\bibliographystyle{plos2015}
\bibliography{LDDMM}
%\printbibliography
\clearpage

\subsection*{Competing Interests}
M.I.M. reports personal fees from AnatomyWorks, LLC, outside the submitted work and jointly owns AnatomyWorks. This arrangement is being managed by the Johns Hopkins University in accordance with its conflict of interest policies. M.I.M.'s relationship with AnatomyWorks is being handled under full disclosure by the Johns Hopkins University. 

\subsection*{Corresponding Author}
Brian C. Lee (blee105@jhu.edu)

\clearpage
\appendix
\title{\Huge \textbf{Supporting Information}}
\section{Reproducing Kernel Hilbert Space and Green's Kernel}
\label{Green-kernel-rkhs}
The Green's kernel is translation invariant and takes the form
$$ K(x, y,z )= k(x, y,z ) \textit{Id}_3 \ , $$ with $\textit{Id}_3$ the $3 \times 3 $ identity matrix, for the Green's function continuously 
differentiable:
$$k(x,y,z) = 4 \left(3 + 3 \sqrt{ x^2+y^2+z^2} + 3  (x^2+y^2+z^2) \right) e^{- \sqrt{ x^2+y^2+z^2}}  
\ .
$$
This Green's function satisfies $(-\nabla^2+1)^4 k(x,y,z)=\delta(x,y,z)$, where $(-\nabla^2+1)^4$ is referred to as $A$.
The reproducing kernel Hilbert space (RKHS) with this Green's kernel corresponds to vector fields satisfying
$$ \| v \|_V^2 = \sum_{i=1}^3 \int_{\R^3} ((-\nabla^2+1)^2 v_i(x,y,z))^2 dxdydz < \infty \ .$$

\section{Geodesics solving Euler-Lagrange Equations}
\label{geodesic-appendix-section}
The explicit equations for geodesics associated to the RKHS norm $\| v \|_V$
and the geodesics satisfy the Euler-Lagrange equations \cite{Miller2002,Miller2006} given by the triple of equations.
\begin{equation}
\label{geodesics-label}
\begin{cases}
\dot \varphi_t = v _t \circ \varphi_t \\
\dot p_t = - (dv_t)^T \circ \phi_t p_t \\
v_t = \int_{\mathbb{R}^3} K(x,\varphi_t(y))p_t(y) dy
\ , \ Av_0 = p_0 \ .
\end{cases}
\end{equation}

To prove the Hamiltonian momentum evolution the second equation $\dot p = - (dv)^T\circ \phi p $
of \eqref{geodesics-label}
for $Av$ a classical function we use the inner product notation $\langle \cdot , \cdot \rangle $ to calculate the Lagrangian: 
$$L(\phi,\dot \phi) =\frac{1}{2} \langle A \dot \phi \circ \phi^{-1},  \dot \phi \circ \phi^{-1} \rangle =
\frac{1}{2} \int_{{\mathbb R}^3} A (\dot \phi \circ \phi^{-1}(x)) \cdot  \dot \phi \circ \phi^{-1}(x) dx \ ,$$ with the variation giving the
Euler-Lagrange equations:
$$
\frac{d}{dt}\underbrace{ \partial_{\dot \phi} L(\phi,\dot \phi) }_{\text{Ham. mom. p}}- \partial_{\phi} L(\phi,\dot \phi) = 0 .
$$
To get the Hamiltonian momentum $ p =\partial_{\dot \phi} L(\phi,\dot \phi) $, we take variation with respect to Lagrangian velocity $ \dot \phi \rightarrow \dot \phi^\epsilon = \dot \phi +\epsilon \delta \dot \phi$ and $\phi \rightarrow \phi + \epsilon \delta \phi$ giving
\begin{eqnarray}
\frac{d}{d\epsilon} {L}(\phi^\epsilon, \dot \phi^\epsilon)|_{\epsilon=0} &=&\frac{d}{d \epsilon} \frac{1}{2}  \langle A( \dot \phi^\epsilon \circ \phi^{-1}) , \dot \phi^\epsilon \circ \phi^{-1} \rangle |_{\epsilon=0}
\nonumber
\\
&=&
\frac{d}{d \epsilon}  \frac{1}{2}\left( \langle A v , \dot \phi^\epsilon \circ \phi^{-1}\rangle 
+  \langle A ( \dot \phi^\epsilon \circ \phi^{ -1}) , v \rangle \right) |_{\epsilon=0}
\nonumber
\end{eqnarray}
Combining gives the
 Hamiltonian momentum :
\begin{eqnarray}
\langle Av, \frac{d}{d \epsilon} (\dot \phi+\epsilon \delta \dot \phi) \circ \phi^{-1} \rangle 
\label{Lagrangian-variation-velocity}
&=& \langle \underbrace{Av \circ \phi | d \phi |}_{ {\partial_{\dot \phi} L} \text{Ham. mom.}} , \delta \dot \phi \rangle \ .
\nonumber
\end{eqnarray}
The variation $ \phi \rightarrow \phi^\epsilon = \phi + \epsilon \delta \phi$ requires the inverse:
$$ (\phi^{-1}+\epsilon \delta \phi^{-1})\circ (\phi + \epsilon \delta \phi) \simeq \id + \epsilon (d \phi^{-1})_{|\phi} \delta \phi + 
 \epsilon \delta \phi^{-1}_{|\phi }$$
which gives first order perturbation 
%$ \phi^{-1} \rightarrow \phi^{\epsilon-1} = \phi^{-1} + \epsilon \delta \phi^{-1}$ with
\begin{equation}
\label{inverse-perturbation}
\delta \phi^{-1} = - (d \phi^{-1}) \delta \phi_{|\phi^{-1}} =
- (d \phi)_{\phi^{-1}}^{-1} \delta \phi_{|\phi^{-1}}
\ .
\end{equation}
Taking a similar variation of the Lagrangian as above but with respect to the Lagrangian velocity gives
\begin{eqnarray}
% \int A v \cdot \frac{d}{d \epsilon} (\dot \phi \circ \phi^{\epsilon -1}) dx|_{\epsilon=0}=
 \langle A v , \frac{d}{d \epsilon} (\dot \phi \circ (\phi^{ -1}-  \epsilon (d \phi)_{|\phi^{-1}}^{-1} \delta \phi_{|\phi^{-1}})) \rangle
&=&
-\langle A v , (dv )(d \phi)_{|\phi^{-1}}(d\phi)_{|\phi^{-1}}^{-1} \delta \phi_{|\phi^{-1}}  \rangle \ 
\nonumber
\\
&=&
-\langle \underbrace{(dv )_{\phi}^T A v \circ \phi |d \phi|}_{{\partial_\phi L} } , \delta \phi \rangle \ 
\end{eqnarray}
The third equation of \eqref{geodesics-label}
follows from $p= Av \circ \phi |d \phi|$.
Integrating with the Green's kernel gives the expression
$v_t(\cdot) = \int K(\cdot, \phi_t (y)) p_t(y) dy $.
\section{Gradients for Atlas Free Model}
\label{Gradients-atlas-free-model-appendix}
We can write the gradient with respect to the components of $R$ (translation vector $t$ and rotation matrix $r$ parametrized by rotation angle $\theta$ and section number $z$), where $\nabla_X$ is the 2D in-plane gradient, $\sigma_{JJ}$ is the weighting factor on the image smoothness prior. Rotations and translations are penalized by a regularization prior centered at identity ($\frac{\theta}{\sigma_{regR}^2}$ and $\frac{t(z)}{\sigma_{regT}^2}$, respectively), where $\sigma_{regR}$ and $\sigma_{regT}$ are weighting factors on the rotation and translation priors.
\begin{multline}
\nabla_r E = - \frac{1}{\sigma_{JJ}^2} \frac{d^2}{dz^2} \left( J(r(\theta,z)x + t(z)) \right) \nabla_{X} J(r(\theta,z)x + t(z))\delta r(\theta,z) x + \\ \frac{\theta}{\sigma_{regR}^2}
\end{multline}
\begin{multline}
\nabla_t E = - \frac{1}{\sigma_{JJ}^2} \frac{d^2}{dz^2} \left( J(r(\theta,z)x + t(z)) \right) r(\theta,z),\nabla_{X} J(r(\theta,z)x + t(z)) + \\  \frac{t(z)}{\sigma_{regT}^2}
\end{multline}
\section{Gradients for Random Orbit Model}
\label{Gradients-random-orbit-appendix}
The minimization of the energy $E_v$ of \eqref{LDDMM-minimization} in terms of the vector field
 is the LDDMM gradient of Beg \cite{Beg2005}:
\begin{multline}
\nabla_{v} E_v(x,y)= \sum_i \int_{\mathbb{R}^2}K(x-x^\prime,y-y^\prime,z-z_i) |D \phi_{t,1} | \left( I \circ \phi_{t1}- I_0\circ \phi_{t}^{-1}) \right. \\ \left. \nabla (I_0 \circ \phi_t^{-1})
(x^\prime,y^\prime,z_i) \right ) dx^\prime dy^\prime  \ .
\end{multline}
\subsection{Variation of the Image Matching Term}
The variation of $\int (I-I_0 \circ \varphi^{-1})^2 dx $ via perturbation $ \varphi \rightarrow \varphi^\epsilon = \phi + \epsilon \delta \varphi$ requires the inverse perturbation
$\delta \phi^{-1} = -(d \varphi)_{\varphi^{-1}}^{-1} \delta \varphi|_{\varphi^{-1}} $, derived in \eqref{inverse-perturbation}
above.
Then we have
\begin{eqnarray}
\frac{d}{d \epsilon}  \int_{\mathbb{R}^3} (I-I_0 \circ \varphi^{\epsilon -1})^2 dx |_{\epsilon = 0} &=& 2\int_X (I-I_0\circ \varphi^{-1}) \nabla (I_0){| \varphi^{-1}}  \cdot 
 (d \varphi)_{|\varphi^{-1}}^{-1} \delta \varphi_{|\varphi^{-1}}) dx
\nonumber
\\
&=& 2\int_X (I \circ \varphi -I_0)  (d \varphi)^{-1 T}\nabla I_0  |d \varphi | \cdot 
 \delta \varphi dx
\ .
\nonumber
\end{eqnarray}

\subsection{Rigid motion variations}
Rigid motion minimization is standard for rigid registration in 2D and 3D images.
Denoting $ \| f_{\theta,t,z_i}\|^2=\| J^R (\cdot,z_i) -  I_0 \circ \varphi^{ v^* -1}(\cdot,z_i)\|_2^2 $ to represent each rigid registration norm-square minimization within each histological plane, then
\begin{eqnarray}
\nabla_\theta \| f_{\theta,t,z_i} \|^2= \int_{\mathbb{R}^2} 2 f_{\theta,t,z_i}(\cdot) \frac{\partial_\theta f_{\theta,t,z_i}}{\partial \theta} dx dy \ ;
\nonumber
\\
\nabla_t \| f_{\theta,t,z_i} \|^2= \int_{\mathbb{R}^2} 2 f_{\theta,t,z_i}(\cdot) \nabla_t f  dx dy \ .
\nonumber
\end{eqnarray}
\begin{multline}
\nabla_{R,t} \ell(v,R;J) = \Big\langle \frac{1}{\sigma_{JI}^2} (I_{\varphi^{-1}}(x) - J(r(\theta,z)x + t(z))) - \frac{1}{\sigma_{JJ}^2} \frac{d^2}{dz^2} \left( J(r(\theta,z)x + t(z)) \right) r(\theta,z),\\\nabla_{X} J(r(\theta,z)x + t(z)) \Big\rangle + \frac{t(z)}{\sigma_{reg_t}^2}
\end{multline}
\begin{multline}
\nabla_{R,r} \ell(v,R;J) = \Big\langle \frac{1}{\sigma_{JI}^2} (I_{\varphi^{-1}}(x) - J(r(\theta,z)x + t(z))) - \frac{1}{\sigma_{JJ}^2} \frac{d^2}{dz^2} \left( J(r(\theta,z)x + t(z)) \right),\\\nabla_{X} J(r(\theta,z)x + t(z))R \left[ \begin{array}{cc} 0 & 1\\-1 & 0\end{array} \right] x \Big\rangle + \frac{\theta}{\sigma_{reg_r}^2}
\end{multline}
where $\sigma_{JI}$ is a weighting factor on the matching term between atlas and target.

\end{document}